\documentclass[12pt]{article}
\input epsf
\usepackage{amsfonts}
\usepackage{amsmath}
\usepackage{amssymb}
\usepackage{bbm}
\usepackage{dsfont}
\usepackage{graphicx}
\usepackage{hyperref}

\newcommand{\bea}{\begin{eqnarray}}
\newcommand{\eea}{\end{eqnarray}}
\newcommand{\be}{\begin{equation}}
\newcommand{\ee}{\end{equation}}
\newcommand{\nn}{\nonumber}
\newcommand{\pa}{\partial}
\newcommand{\tr}{\mbox{tr}}
\newcommand{\Tr}{\mbox{Tr}}

\newcommand{\Z}{{\mathbb Z}}
\newcommand{\C}{{\mathbb C}}

\newcommand{\D}{\Delta^0}
\newcommand{\ttau}{\tilde{\tau}}
\newcommand{\ttheta}{\tilde{\theta}}
\newcommand{\tc}{\tilde{c}}
\newcommand{\tV}{\tilde{V}}

\renewcommand{\Re}{\mbox{Re}}
\renewcommand{\Im}{\mbox{Im}}

\newcommand{\A}{{\cal A}}

\makeatletter

\@addtoreset{equation}{section}
\makeatother

\def\href#1#2{#2}

\begin{document}

\begin{titlepage}

\begin{center}

\hfill 
\vskip 0.8 in

{\bf \Large Gauge-Higgs Unification In}\\[3mm]
{\bf \Large Spontaneously Created Fuzzy Extra Dimensions}\\[18mm]

\centerline{Kazuyuki Furuuchi${}^1$,
Takeo Inami${}^2$ and Kazumi Okuyama${}^3$}
\vskip10mm
\centerline{${}^1$\it National Center for Theoretical Sciences}
\centerline{\it National Tsing-Hua University, Hsinchu 30013, Taiwan,
R.O.C.}
\centerline{\tt furuuchi@phys.cts.nthu.edu.tw}
\vskip2mm
\centerline{${}^2$\it Department of Physics, Chuo University}
\centerline{\it Bunkyo-ku, Tokyo 112, Japan}
\centerline{\tt inami@phys.chuo-u.ac.jp}
\vskip2mm
\centerline{${}^3$\it Department of Physics, Shinshu University}
\centerline{\it Matsumoto 390-8621, Japan}
\centerline{\tt kazumi@azusa.shinshu-u.ac.jp}
\vskip6mm 
\end{center}
\begin{abstract}
We propose gauge-Higgs unification in fuzzy extra dimensions
as a possible solution to the  
Higgs naturalness problem.
In our approach, the fuzzy extra dimensions are created spontaneously as a
vacuum solution of certain four-dimensional gauge theory. 
As an example, we construct a model which has a fuzzy torus as its vacuum. 
The Higgs field in our model is associated with
the Wilson loop wrapped on the fuzzy torus. 
We show that 
the quadratic divergence in 
the mass of the Higgs field
in the one-loop effective potential is absent. 
We then argue based on symmetries that 
the quantum corrections to the Higgs mass is suppressed
including all loop contributions.
We also consider a realization 
on the worldvolume theory of D3-branes probing $\C^3/(\Z_N \times \Z_N)$ 
orbifold with discrete torsion.
\end{abstract}

\end{titlepage}

\tableofcontents

\section{Introduction}

The Standard Model has a naturalness problem
regarding the mass of the Higgs field.
The leading quantum corrections to the Higgs mass square $\delta m_H^2$ 
takes the form
\bea
 \label{QD}
\delta m_H^2 \sim 
\kappa \,
\Lambda_{SM}^2 ,
\eea
where $\kappa$ is a numerical coefficient
and $\Lambda_{SM}$ is the UV cut-off for the Standard Model.
$\Lambda_{SM}$ should be regarded as
a physical energy scale above which
the modification to the Standard Model becomes significant.
The quadratic dependence on the UV cut-off $\Lambda_{SM}$
in (\ref{QD})
is a generic feature of the quantum correction 
to the scalar mass in four dimensional space-time
and this UV sensitivity 
is the origin of the Higgs naturalness problem.
The Standard Model contributions to
$\kappa$ is of order $\sim 10^{-2}$. 
Since the Higgs mass is expected to be in the order of $\sim$ $10^2$ GeV,
the formula (\ref{QD}) tells us that
the mass of the Higgs field requires unnatural fine-tuning
if the UV cut-off $\Lambda_{SM}$ of the Standard Model
goes too much beyond the TeV scale,
not to mention the GUT scale or Planck scale.
In order for the Standard Model to remain natural,
new physics 
must enter at a few TeV scale
to modify the high energy behavior of the Standard Model.
This is one of the main reasons why
the Large Hadron Collider is likely 
to find not only the Higgs particle
but also the new physics beyond the Standard Model.

The new physics relevant for solving 
the Higgs naturalness problem
must replace (\ref{QD}) by 
\bea
 \label{NPQD}
\delta m_H^2 \sim \kappa_{NP}\, \Lambda_{NP}^2 ,
\eea
with small enough\footnote{%
It is zero in some models, e.g. those
based on supersymmetry.} 
coefficient $\kappa_{NP}$,
where $\Lambda_{NP}$ is a UV cut-off of the model
that describes the new physics
if the cut-off $\Lambda_{NP}$, which
is supposed to be hierarchically higher than the TeV scale.
From the effective field theory point of view,
small parameters in a theory,
by which $\kappa_{NP}$ is made small,
must be associated with 
a (weakly broken) symmetry
\cite{tHooft}
in order for the model to be natural.
Thus 
solving the Higgs naturalness problem
in the framework of effective field theory
boils down to
identifying the relevant symmetry.\footnote{%
There are two other possible explanations to the smallness of the Higgs mass:
1.~The fundamental scale is at some TeV scale.
2.~The Higgs mass is fine-tuned, probably by the anthropic principle.
The first gives a very strong restriction to 
the possible fundamental theory at the highest energy scale
while the second is out of the framework of effective field theory. 
It is also hard to be very convincing.
We will not pursue these possibilities in this paper.}
While (\ref{QD}) and (\ref{NPQD}) may look similar in the form,
the limitation of the Standard Model was 
that it does not have any symmetry
which can protect the Higgs mass 
from the quantum corrections.\footnote{%
We refer to an amusing essay \cite{Giudice:2008bi}
for the historical background
and relevant references regarding
the Higgs naturalness problem.}

Local gauge symmetry is 
relevant up to the electroweak scale,
and it is expected to be important 
even at much higher energy scale,
as employed in candidates of the fundamental theory like string theory.
It also forbids the mass term of the spin-one particles,
and it is 
a vital candidate 
as a solution to the 
Higgs naturalness problem.
In gauge-Higgs unification scenario, 
the Higgs field is
the zero-mode 
of an extra-dimensional component
of a gauge field in higher dimensions.
It has been shown that the
one-loop correction to the mass of the Higgs in this scenario
is indeed insensitive to the UV cut-off
\cite{Hosotani:1983xw,Hosotani:1988bm,Davies:1988wt,Hatanaka:1998yp}.
The Higgs field
can be associated with the Wilson loop
wrapped around a cycle in the extra dimensions.

Gauge theories in higher dimensions
are 
non-renormalizable and inevitably effective field theories
with a finite UV cut-off.
This itself is not an immediate problem,
though obviously another UV theory
is required above 
the cut-off scale.
Another issue is that
the extra dimensions are given a priori
in higher-dimensional gauge theories.
The extra dimensional space is supposed to be 
determined by the dynamics of some gravitational theory in higher dimensions,
which lies beyond the energy scale described by the gauge theory.
It will be interesting if there is 
an alternative scenario 
based on four-dimensional quantum field theory 
where the extra dimensions effectively emerge:
Four-dimensional quantum field theories have more chances to be renormalizable,
and the emergent extra dimensions are described within the framework of
the four-dimensional quantum field theory.
The idea of (de)construction 
\cite{ArkaniHamed:2001ca,ArkaniHamed:2001nc} 
(see also \cite{Hill:2000mu})
realizes such idea
using quiver gauge theory.
In this scenario, the quiver diagram (moose)
prepares latticized extra dimensions,
while the lattice spacing is dynamically determined
by the four-dimensional quiver gauge theory.

On the other hand,
fuzzy spaces are ubiquitous in 
multiple D-brane systems in string theory
\cite{Alekseev:1999bs,Seiberg:1999vs,%
Aoki:1999vr,Myers:1999ps}.
In fact 
it has been known even before the (de)construction 
that the fuzzy extra dimensions can be described as
a vacuum of lower dimensional quantum field theories.
Moreover, it has also been shown that 
the fluctuations around the fuzzy vacuum
contain excitations that can be identified
with a gauge field on the fuzzy space.
Thus fuzzy extra dimensions in string theory
appear as a rather natural 
setting for the 
four-dimensional models of gauge-Higgs unification.
Indeed, this possibility has been noticed for a while, 
see e.g.
\cite{Chatzistavrakidis:2010tq} and references therein.
However, as far as we have noticed, there has been no
detailed study of the quantum aspects of the 
gauge-Higgs unification in 
fuzzy extra dimensions
which is relevant for the Higgs naturalness problem.
The purpose of this work is to construct an explicit model 
that realizes the gauge-Higgs unification
in emergent fuzzy extra dimensions, 
and study its quantum aspects in detail
to make clear 
the issues in this scenario
in the context of Higgs naturalness problem.\footnote{%
Somewhat different quantum aspects
of dynamically generated fuzzy shpere
have been studied in \cite{Aschieri:2006uw}.}
We will be particularly interested in 
the fuzzy extra dimensions
realized by finite size matrices.
In this case,
the KK mass spectrum 
in the fuzzy extra dimensions are truncated
at finite level, 
and the difference
from the ordinary extra dimensions becomes sharp.

%
%
%

\section{A unitary matrix model of gauge-Higgs unification 
in fuzzy torus}\label{secum}

In this section, we study a model
which realizes the gauge-Higgs unification
in fuzzy torus extra dimensions.
At this stage, our model is a toy model
and the ``Higgs" field here means 
a
scalar field in four dimensions and
in some representation of a
gauge group. 
The gauge group should contain
the electroweak gauge group as a subgroup 
and the Higgs field should be in a certain representation
in a realistic model,
but we will not be concerned with these points too much in the following.
This issue has been studied extensively 
in the ordinary gauge-Higgs unification models,
and we leave this issue to the Discussion section.

In the study of
the ordinary gauge-Higgs unification models,
the torus has been
a nice example of the extra dimensions 
in which one could make detailed studies
as well as construct realistic models.\footnote{%
Circle may be the simplest extra dimension,
but fuzzy spaces should have coordinates
which do not commute with each other,
thus we need at least two dimensions.}
Therefore, it would be a good starting point
to study 
a model which has a fuzzy version of the torus as its vacuum.
A brief summary of fuzzy torus is provided in 
the appendix \ref{AfuzzT}.

\subsection{The unitary matrix model}

Let us consider the following
four-dimensional 
action with
$SU(k N)$ gauge group:
\bea
\label{action}
S
&=&
\int d^4 x\,
\tr_{SU(kN)} 
\Biggl[
-\frac{1}{2} F_{\mu\nu}(x) F^{\mu\nu} (x)
+
\sum_{I=1,2}
f_I^2
D_{\mu} U_I (x) D^{\mu} U_I^\dagger (x)\nn \\
&&\qquad \qquad \qquad \qquad \qquad 
+\,
c_0  
U_1 U_2 U_1^{\dagger} U_2^{\dagger}
+
c_0^* 
U_2 U_1 U_2^{\dagger} U_1^{\dagger}
+ \ldots
\Biggr]
\nn \\
&& 
\qquad \qquad \qquad
\qquad (\mu,\nu = 0,\cdots,3), 
\eea
We regard the action (\ref{action})
as an effective field theory with a 
UV cut-off $\Lambda$. 
Like in the chiral perturbation theory,
a natural UV cut-off scale 
may be the energy scale where the 
perturbative loop expansion of the model breaks down
\cite{Weinberg:1978kz,Manohar:1983md}.
As explained in the appendix \ref{estimate},
it is estimated to be 
\bea
 \label{Lambda}
\Lambda \approx \frac{4 \pi f}{\sqrt{kN}} .
\eea
Here, we consider the case
where there is no big hierarchy between the 
scales $f_1$ and $f_2$: $f \approx f_1 \approx f_2 $.
The action (\ref{action}) has two small expansion parameters
and `` $\ldots$ '' in (\ref{action}) denotes the terms
suppressed by powers of these small parameters:
One 
is  
the inverse of the cut-off $1/\Lambda$,
as is usual in effective field theory.
Another 
is a small dimensionless 
$SU(kN)$ gauge coupling $g$,
which in the perturbative expansion
appears in the combination (see the appendix \ref{estimate})
\bea
 \label{gexp}
\frac{gf}{\Lambda} .
\eea
The parameter
(\ref{gexp}) is associated with a breaking of a ``chiral" symmetry
in the action (\ref{action}), as will be explained 
in more detail in section \ref{secorder}.

The fields
$U_I$ 
take values in special unitary matrix:
\bea
U_I^{\dagger}(x) = U_I^{-1} (x), \quad \det U_I (x) =1
\quad (I=1,2).
\eea
The field strength of the gauge field is given as usual:
\bea
F_{\mu\nu}(x) 
= \pa_\mu A_\nu(x) -\pa_\nu A_\mu (x) + i g [A_\mu (x), A_\nu (x)] .
\eea
The covariant derivatives 
are given by
\bea
 \label{CD}
D_{\mu} U_I (x) = \pa_\mu U_I (x) - i g [A_{\mu}(x), U_I (x) ]  .
\eea
The potential part of the action (\ref{action})
has the same form
with the finite rank version of the twisted Eguchi-Kawai model
of lattice gauge theory
\cite{GonzalezArroyo:1982hz,GonzalezArroyo:1982ub}.
In the twisted Eguchi-Kawai model,
the fields
$U_I$ 
are the link fields of the lattice gauge theory
in the extra dimensions, where
two extra dimensions are 
periodic lattice with just one lattice point.
In the twisted Eguchi-Kawai model,
larger size 
extra dimensions are
effectively generated by the vacuum configuration,
as we explain below.\footnote{%
The Eguchi-Kawai 
reduction \cite{Eguchi:1982nm}
may be one of the earliest examples where 
the space(-time) effectively emerges
from a lower dimensional quantum field theory.}

The action (\ref{action}) is an extreme version
of the one 
considered in (de)construction \cite{ArkaniHamed:2001nc}.
In (de)construction, 
latticized extra dimensions
are constructed from the quiver diagram (moose)
of a quiver gauge theory.
In the language of the quiver gauge theory,
our moose has only one node.
The new ingredient of our model 
is that the large 
extra dimensions are generated not by the large moose
but by the fuzzy torus vacuum.
One may regard that the moose effectively gets large
via (the inverse of) the twisted Eguchi-Kawai reduction.

From the point of view of effective field theory,
there is a natural magnitude for the coefficient $c_0$
appearing in the action (\ref{action}).
It is estimated in the appendix \ref{estimate}
and can be parametrized as
\bea
 \label{tczero}
c_0 = g^2 f_1^2 f_2^2\, \tc_0 ,
\eea
where $\tc_0$ is a dimensionless complex number
of order one.

As an effective field theory,
it is important to specify
the symmetries the action (\ref{action}) has. 
We impose the four-dimensional
Poincare symmetry and 
the $SU(kN)$ gauge symmetry
as exact symmetries. 
We also require the action 
to be invariant under the following 
${\mathbb Z}_{kN} \times {\mathbb Z}_{kN}$
global transformations:
\bea
 \label{ZNk}
U_I \rightarrow e^{\frac{2\pi i}{kN} n_I} \, U_I  \quad
(n_I \in {\mathbb Z}_{kN},\, I = 1,2).
\eea
This is the 
so-called
center symmetry 
which often appears
in the study of gauge theories
with $SU(kN)$ gauge group.
It is particularly important in the Eguchi-Kawai reduction
since the condition for the Eguchi-Kawai reduction to take place
is that this symmetry (or the large part of it, see below)
is not broken.
The $\Z_N \times \Z_N$ subgroup of 
the $\Z_{kN} \times {\mathbb Z}_{kN}$ global symmetry (\ref{ZNk})
will be crucial for the suppression
of the quantum corrections to the mass of 
our model Higgs field, 
as will be discussed
in section \ref{secorder}.

In effective field theories,
not only the exact symmetries
but also approximate 
symmetries play important roles.
Let us consider the CP transformation:
\bea
 \label{CP}
A_\mu \rightarrow A_\mu^T , \quad 
U_I \rightarrow U_I^T .
\eea
CP symmetry is broken by the 
following term in the action (\ref{action}):
\bea
 \label{odd}
i\, \Im \, c_0 \, \tr_{SU(kN)}
\left[
U_1 U_2 U_1^{\dagger} U_2^{\dagger}
-
U_2 U_1 U_2^{\dagger} U_1^{\dagger}
\right].
\eea
This means that
it is natural for the coefficient
$\Im \, \tc_0$
to be small in the sense of 't Hooft \cite{tHooft}.

The following transformations 
which can be regarded as the
reflections of coordinates
in the extra dimensional lattice directions
are also weakly broken by the term (\ref{odd}):
\bea
 \label{reflect1}
P_1 &:& U_1 \rightarrow U_1^{-1} = U_1^\dagger,
\\
P_2 &:& U_2 \rightarrow U_2^{-1} = U_2^\dagger
\label{reflect2}.
\eea 

In addition to the symmetries mentioned above,
the leading terms presented in the 
action (\ref{action}) has a weakly broken
global $(SU_L(kN)\times SU_R(kN) )^2$
``chiral" symmetry\footnote{%
The reason we call it chiral symmetry is that
chiral gauge theories are candidates
of the UV completion of this effective field theory,
and the origin of this symmetry in this case is the
approximate chiral symmetry \cite{ArkaniHamed:2001nc}.}
which recovers 
when the gauge coupling $g$ is turned off:\footnote{%
We have taken into account (\ref{tczero}).}
\bea
 \label{chiral}
U_I \rightarrow L_I U_I R_I^{\dagger}
\qquad (I = 1,2),
\eea
where $L_I$ and $R_I$ are independent 
$SU(kN)$ matrices.
A subgroup of this chiral symmetry
is the origin of the small expansion parameter
(\ref{gexp})
and will be 
crucial for the suppression
of 
the quantum corrections to the mass of 
the Higgs field, as we discuss in section \ref{secorder}.

The potential term in the action (\ref{action})
which is leading in the expansions in 
$1/ \Lambda$ and $fg/\Lambda$ is
\bea
 \label{pot}
V_0(U_I) 
\equiv
- 
\tr_{SU(kN)}
\left[
c_0 
U_1 U_2 U_1^{\dagger} U_2^{\dagger}
+
c_0^* 
U_2 U_1 U_2^{\dagger} U_1^{\dagger}
\right].
\eea
This can be rewritten in the form of the perfect square:
\bea
 \label{pot2}
V_0(U_I) 
= 
g^2 f_1^2 f_2^2 | \tc_0 |\,
\tr_{SU(kN)}
\left[
\left| 
U_1 U_2 - e^{-i\theta} U_2 U_1
\right|^2
-2
\right],
\eea
where 
we have used the parametrization (\ref{tczero}) and
$\theta$ is the phase of the complex number $\tc_0$:
\bea
\tc_0 = |\tc_0| e^{i\theta} .
\eea
Then, 
the absolute minimum of the potential (\ref{pot})
is given by the configuration satisfying
\bea
 \label{FTeq}
U_1 U_2 - e^{-i\theta} U_2 U_1 = 0.
\eea
However, (\ref{FTeq}) is satisfied only for
specific values of $\theta$,
as we describe shortly. 

We would like to consider the vacuum configuration of the form
\bea
 \label{UV}
&&U_1 = V_1 \equiv W_1 \otimes e^{\frac{N-1}{N} \pi i} \mathds{1}_k , \nn \\
&&U_2 = V_2 \equiv W_2 \otimes e^{\frac{1}{N} \pi i} \mathds{1}_k ,
\eea
where $W_1$ and $W_2$ are $N \times N$
constant unitary matrices satisfying
the relation
\bea
 \label{ZZ}
W_1 W_2 = e^{-i \theta} W_2 W_1 .
\eea
The phases in (\ref{UV})
are put to make the fields $U_I$ 
to be special unitary matrices.
Notice that
in order to satisfy (\ref{ZZ})
by finite size matrices,
the parameter $\theta$
has to take a special value
\bea
 \label{theta}
\theta = \frac{2\pi}{N} \, \ell ,
\eea
where $\ell$ is an integer. 
This can be understood by taking the determinant of
both sides of (\ref{ZZ}).
We will discuss the case when 
$\theta$ is away from
the value (\ref{theta})
in section \ref{dynth}.
Our purpose here is to explain
the mechanism
that suppresses 
the quantum corrections to the Higgs mass.
Therefore we may 
choose the simplest case $\ell =1$ 
as an example.\footnote{If $\ell$ is a divisor of $N$,
we can redefine 
$N_{new} = N/\ell$,
$k_{new} = \ell k$ to have $\ell_{new} = 1$.
If we assume that the CP violation due to this phase is small,
$\ell/N$ is naturally small.
Put it differently, the model with 
$SU(N')$ gauge group ($N' = kN$)
can have fuzzy torus solution with $\theta = 2\pi \ell / N'$,
and we chose the $\ell=k$ case with $k$ being a divisor of $N'$.}
The case for other $\ell$ is similar
as long as $\ell$ is small compared with $N$.

The vacuum (\ref{UV}) breaks the gauge symmetry to 
$SU(k)$.
In a realistic model, the electroweak gauge group 
should be in a subgroup of this $SU(k)$ gauge group.
The vacuum (\ref{UV}) also breaks the global ${\mathbb Z}_{kN} \times \Z_{kN}$
symmetry to ${\mathbb Z}_{N} \times {\mathbb Z}_{N}$.
The global ${\mathbb Z}_{N} \times {\mathbb Z}_{N}$ symmetry
\bea
 \label{Zn}
U_I \rightarrow e^{\frac{2\pi i}{N} n_I} \, U_I 
\quad (n_I \in {\mathbb Z}_N, I = 1,2),
\eea
is not broken since the ${\mathbb Z}_{N} \times {\mathbb Z}_{N}$ 
transformation (\ref{Zn}) 
to the vacuum (\ref{UV}) is 
equivalent to a
gauge transformation due to (\ref{UV}):
\bea
 \label{Zngauge}
e^{\frac{2\pi i}{N}} V_1 &=& V_2 V_1 V_2^{\dagger} , \nn\\
e^{\frac{2\pi i}{N}} V_2 &=& V_1^{\dagger} V_2 V_1 .
\eea
This unbroken $\Z_N \times \Z_N$ symmetry
will be crucial
for the suppression of the quantum corrections
to the mass of the Higgs field.
This will be explained in section \ref{secorder}.

The configuration (\ref{UV})
can be interpreted as a fuzzy torus \cite{Ishibashi:1999hs,Ambjorn:1999ts}.
The reason that 
it can be regarded as a fuzzy version of the {\it torus} is that
the mass spectrum on this vacuum approximates
the low-lying KK modes on the ordinary torus,
as we will see below. 
On the other hand, 
the higher mass spectrum
deviates from that of the KK modes on the ordinary torus
and is truncated at a finite level.\footnote{%
When the size $N$ of the matrices which
describe the fuzzy torus is finite. 
In this paper, we will mostly consider 
this case.
See section \ref{secorder} for further discussions.}
Thus one cannot probe arbitrarily small distance
on the fuzzy torus, and the notion of a point
becomes obscure.
This is the reason we call it {\it fuzzy}.

\subsection{One-loop effective potential around
the fuzzy torus vacuum}

For concreteness, below we study the case $k=2$.
Generalization to arbitrary $k$ is straightforward. 
In the ordinary gauge-Higgs unification scenario,
the Higgs field is identified with a zero-mode of
an extra dimensional component of a gauge field.
We identify corresponding zero-modes
in 
our fuzzy torus model below.
Then 
we 
calculate the
1PI effective potential for the
zero-modes. 

The fluctuations of the field $U_I$
around the fuzzy torus vacuum (\ref{UV})
are analogous to the link variables
in the lattice gauge theory
(see the appendix \ref{AfuzzT}).
Thus these fluctuations
can be identified with 
the exponential of
the components of the gauge field
in the extra dimensions.
On the other hand,
the commutators with the vacuum configuration $V_I$ (\ref{UV})
are related to the discrete counterparts of the
derivatives on the fuzzy torus
(see the appendix \ref{AfuzzT}).
Thus the zero-modes in the extra dimensions are
those which commute with the matrices $V_I$.
Without loss of generality, we can parametrize the zero-modes 
by $u_0^I(x)$ as
\bea
\label{Uo}
U_I = U_{I}^{(0)} \equiv  e^{i 
\frac{u_0^I(x)}{\sqrt{4N} f_I}  \Sigma }
V_I
\qquad (I=1,2),
\eea
where
\bea
\Sigma = \mathds{1}_N \otimes  \sigma_3 ,
\eea
and $\sigma_i$ $(i=1,2,3)$ are the Pauli matrices.
In (\ref{Uo})
we have canonically normalized the
zero-modes $u_0^I(x)$. 
Notice that since the zero-modes 
$u_0^I(x)$ 
commute with the matrices $V_I$
as well as with each other by definition,
$U^{(0)}_{I}$ also satisfy the relation
\bea
U_{1}^{(0)} U_{2}^{(0)} = e^{-i\theta} U_{2}^{(0)} U_{1}^{(0)}.
\eea
Thus the configuration $U_I^{(0)}$ is also a classical minimum
of the leading potential (\ref{pot2}).
In other words, the zero-modes $u_0^I$ parametrize
the classical flat directions of (\ref{pot2}).

Now we calculate 
the 1-PI effective potential 
for $u_0^I(x)$ from 
the one-loop diagrams made from the leading terms
explicitly shown in the action (\ref{action}).\footnote{%
In the current choice of the UV cut-off $\Lambda$ (\ref{Lambda}), 
the natural magnitude of a term at the tree level
is the same to that of the loop contributions.}
For simplicity, we present the calculation for $\tc_0 = 1$.
When $\tc_0 \ne 1$,
the gauge field $A_\mu$ and the field $u^I$ introduced below
feel different fuzzy torus radii.
Because of this the calculation for $\tc_0 \ne 1$ case
is slightly more complicated compared with the $\tc_0 = 1$ case.
However, 
$\tc_0 = 1$ case is enough for understanding
the mechanism that suppresses
the quantum corrections to the Higgs mass.
We will also give 
more general analysis for the suppression of the quantum corrections
based on symmetries in section \ref{secorder}. 

We fix the gauge as
\bea
 \label{gfix}
\pa_\mu A^\mu + 
\D_I u^I = 0,
\eea
where we have parametrized $U_I$ as
\bea
U_I = e^{i \frac{u^I(x)}{\sqrt{4N} f_I}} U_I^{(0)}  .
\eea
We have also defined
\bea
 \label{difference}
\D_I \varphi \equiv \frac{1}{a_I} 
\left( U_I^{(0)} \varphi U_I^{(0)\dagger} - \varphi \right),
\eea
for a field $\varphi$ in the adjoint representation
of $SU(kN)$, where 
\bea
 \label{agf}
a_I \equiv \frac{1}{g f_I}\quad (I =1,2).
\eea
Since we have assumed $f_1 \approx f_2 \approx f$,
$a_1 \approx a_2 \approx a$.
(\ref{difference})
are discrete counterparts of
the covariant derivatives
with 
the background gauge field.
Including the contributions
from the ghost fields, 
the one-loop effective potential is given as 
\bea
 \label{Veff}
V_{1-loop} (u_0^I)
&=&
i \ln \det ((D^0)^2)^{-6/2} + i \ln \det ((D^0)^2)^{+1}
\nn \\
&=&
- 2i\, \Tr \ln ((D^0)^2) ,
\eea
where we have defined
\bea
(D^0)^2 \equiv \pa_\mu \pa^\mu + \D_I \D_I  .
\eea
After the Wick rotation
(\ref{Veff}) becomes
\bea
 \label{Veff2}
V_{1-loop} (u_0^I)
&=&
2 
\sum_{m_1,m_2} 
\sum_{i,j=1}^2
\int \frac{d^4 k}{(2\pi)^4} 
\ln \left(k^2 + m^2_{(m_1,m_2)(i,j)}(u_0^I) \right) ,
\eea
where
\bea
 \label{mass}
m^2_{(m_1,m_2)(i,j)} (u_0^I) 
&\equiv&
\sum_{I =1,2}
\left(\frac{2}{a_I}\right)^2
\sin^2 \frac{1}{2} \left( m_I \theta + (u_i^I - u_j^I) \right) 
\nn \\
&=&
\sum_{I=1,2}
\frac{2}{a_I^2}
\left(1 - \cos \left( m_I \theta + (u_i^I - u_j^I) \right) \right) ,
\eea
with
\bea
u_1^I = - u_2^I =  
\frac{1}{\sqrt{4N} f_I} u_0^I .
\eea
In (\ref{Veff}) the sum over
$m_I$ $(I=1,2)$ run 
over integers in 
$ -\frac{N}{2} \leq m_I < \frac{N}{2}$.
Notice that when $m_I \ll N$,
the mass spectrum
of the fluctuations around the 
fuzzy torus vacuum (\ref{UV})
($u_0^I(x)=0$ in (\ref{mass})) 
approximates the low-lying KK modes 
of the ordinary torus 
\bea
\frac{2}{a_I^2} 
\left( 1 - \cos ( m_I \theta ) \right)
\approx
\left( \frac{m_I}{R_I} \right)^2 ,
\eea
where the radii $R_I$ of the torus are given by
\bea
{2\pi}R_I \equiv {N a_I} \qquad (I=1,2).
\eea
This is the reason why
the vacuum (\ref{UV})
is regarded as the fuzzy version of the torus.
Below we will call $R_I = a_I N / 2\pi$ as the radii of the fuzzy torus.

With the
momentum UV cut-off at $\Lambda$,
(\ref{Veff2}) becomes
\bea
\label{oneloop}
V_{1-loop} (u_0^I)
&=&
\sum_{m_1,m_2} \sum_{i,j=1}^2
\Biggl[
\frac{\Lambda^2}{8\pi^2} m_{(m_1,m_2)(i,j)}^2(u_0^I)  \nn \\
&& + \frac{1}{16 \pi^2} (m_{(m_1,m_2)(i,j)}^2(u_0^I))^2 
\ln \frac{m_{(m_1,m_2)(i,j)}^2(u_0^I)}{\Lambda^2}
+ {\cal O}(\Lambda^{-2}) \Biggr] .
\eea
Now, notice that 
the sum of $m^2_{(m_1,m_2)(i,j)} (u_0^I)$ over $m_1$ and $m_2$
does not depend on $u_0^I(x)$ for $N \geq 2$.
Similarly, the sum of $(m^2_{(m_1,m_2)(i,j)} (u_0^I) )^2$ over $m_1$ and $m_2$
does not depend on $u_0^I(x)$ for $N \geq 3$,
due to the cancellations between phases. 
Recall that we are considering the case $\ell=1$ in (\ref{theta})),
i.e. $\theta = 2\pi / N$.
Thus the divergences
associated with
$\Lambda \rightarrow \infty$\footnote{\label{footdiv}%
Below we will call divergences 
associated with taking the cut-off $\Lambda$ to infinity as
``divergence" for briefness,
although the natural UV cut-off scale 
is as in (\ref{Lambda}) in our model.}
contribute only to the constant term in the 
effective potential for the zero modes $u_0^I(x)$ for $N \geq 3$,
while there remains only logarithmic divergences
for $N = 2$.\footnote{%
This result is similar to that in
the (de)construction models \cite{ArkaniHamed:2001nc}.}

Similarly, we observe that
for given $N$
the first non-zero correction
in the inverse power expansion of the UV cut-off $\Lambda$
is proportional to
\bea
\label{divstr}
\sum_{m_1,m_2}\sum_{i,j=1}^2
\Lambda^4
\left(
\frac{m_{(m_1,m_2)(i,j)}^2(u_0^I)}{\Lambda^2}
\right)^{N} .
\eea
We will discuss this structure
from the point of view of effective field theory below.

The mass $m_0$ of the zero-modes $u_0^I$ 
in the one-loop effective potential (\ref{oneloop})
is calculated in the appendix \ref{Aoneloop}
and of the order
\bea
m_0^2
\approx
\frac{g_{SU(k)}^2}{16\pi^2}
\frac{1}{R^2} ,
\eea
where the four-dimensional $SU(k)$ gauge coupling 
$g_{SU(k)}$ is given in (\ref{gscale}) 
and
$R_1 \approx R_2 \approx R$ follows from our earlier
assumption $f_1 \approx f_2 \approx f$.
This is as expected since
$1/R$ is the scale where 
the effect of the new physics appears,
and
$g_{SU(k)}^2 / {16\pi^2}$
is the one-loop factor.

To generalize the above $k=2$ result to 
general $k$,
just notice that the zero-modes are parametrized
by the Cartan of the $SU(k)$ group.
The result has the same form with
(\ref{oneloop}),
with the sum over the indices $i$ and $j$ run from $1$ to $k$.

\subsection{Fuzzy torus local minima for general $\theta$}\label{dynth}

So far we have been considering a special case
where the parameter $\theta$ takes the particular discrete value
(\ref{theta}).
Below we will argue that 
our previous analysis can be applied to
the case when $\theta$ is away from this value,
with just minor modifications.

We would like to analyze the minima of the effective potential\footnote{%
Rather than the value of $\theta$ in the action, 
the corresponding parameter
in the 1-PI effective potential
is directly relevant for the determination of the vacuum.
Below we use the same symbols $\theta$, $\tc_0$ etc. 
to express the parameters 
in the 1-PI effective potential.}
\bea
 \label{potth}
V_0(U_I;\theta) = g^2 f_1^2 f_2^2 |\tc_0| \tV_0 (U_I;\theta) ,
\eea
where
\bea
 \label{tpotth}
\tV_0 (U_I;\theta)
\equiv 
-
\tr_{SU(kN)} 
\left[
e^{i \theta}
U_1 U_2 U_1^\dagger U_2^\dagger
+
e^{-i \theta}
U_2 U_1 U_2^\dagger U_1^\dagger
\right].
\eea
(\ref{potth}) is the leading term of the effective potential in
the expansions in $1/\Lambda$ and $gf/\Lambda$.

Let us define 
fuzzy torus background $V_I(\ell)$ for $\theta_\ell$
as follows:
\bea
\theta_\ell \equiv \frac{2\pi}{N} \, \ell \quad (\ell: \mbox{integer}),
\eea
\bea
 \label{FTl}
V_1(\ell) V_2(\ell) = e^{- i \theta_\ell} V_2(\ell) V_1(\ell).
\eea
Let us put $U_I = V_I(\ell)$ into the potential (\ref{potth})
and study the fluctuations around the background.
We can rewrite (\ref{tpotth}) as
\bea
\tV_0 (U_I;\theta)
&=& 
- \cos \delta \theta \,
\tr_{SU(kN)} 
\left[
e^{i \theta_\ell}
U_1 U_2 U_1^\dagger U_2^\dagger
+
e^{-i \theta_\ell}
U_2 U_1 U_2^\dagger U_1^\dagger
\right] 
 \label{tpot1} \\
&& 
- i \sin \delta \theta \,
\tr_{SU(kN)} 
\left[
e^{i \theta_\ell}
U_1 U_2 U_1^\dagger U_2^\dagger
-
e^{-i \theta_\ell}
U_2 U_1 U_2^\dagger U_1^\dagger
\right],  
\label{tpot2} 
\eea
where
\bea
\delta \theta \equiv \theta - \theta_\ell .
\eea
The fluctuation spectrum around the $U_I = V_I(\ell)$
in the first line (\ref{tpot1})
can be analyzed similarly as in the case $\ell=1$.
When $\cos \delta \theta > 0$,
the fuzzy torus vacuum is a local minimum of the term (\ref{tpot1})
and there are massless and massive fluctuations but no
tachyonic modes.
On the other hand,
if we put $U_I = V_I(\ell)$ into (\ref{tpot2}) and
expand $U_I$ around this background,
one can explicitly check that
the terms linear or quadratic in the fluctuations are absent.
This means that the fuzzy torus configuration
stays in the local minimum of the potential (\ref{tpotth})
as long as $\cos \delta \theta > 0$.

We can  understand the absence of 
the linear and the quadratic fluctuations
around the fuzzy torus in the term (\ref{tpot2})
from the symmetry.
Consider the 
CP transformation (\ref{CP}):
\bea
 \label{CPth}
&&A_\mu \rightarrow A_\mu^T , \quad 
U_I \rightarrow U_I^T , \nn\\
&&\theta \rightarrow -\theta .
\eea
Here, we have associated a transformation property to 
the parameter $\theta$ 
under the CP transformation
so that the whole action becomes symmetric under the
CP transformations.
Under the CP transformation,
both the term
\bea
\tr_{SU(kN)} 
\left[
e^{i \theta_\ell}
U_1 U_2 U_1^\dagger U_2^\dagger
-
e^{-i \theta_\ell}
U_2 U_1 U_2^\dagger U_1^\dagger
\right],
\eea
and the coefficient $i \sin \delta \theta$
are odd, so that the whole potential is even under the CP transformation.
Here, the CP transformation 
property of $\theta_\ell$ and $\delta \theta$ are
induced from that of $\theta$:
$\theta_\ell \rightarrow - \theta_\ell$,
$\delta \theta \rightarrow - \delta \theta$.
On the other hand,
From the CP symmetry 
and the gauge symmetry,
the only possible terms
linear and quadratic in the fluctuations
corresponds to the following term in the continuum limit $N \rightarrow \infty$:
\bea
 \label{topological}
i \delta \theta \int_{T^2} \tr_{SU(k)} \, F_{12},
\eea
($F_{12}$ is the field strength of the $SU(k)$ gauge theory
on the torus)
which is identically zero 
(it would be a total derivative if we were considering 
the $U(k)$ gauge group instead of $SU(k)$,
which is again zero after the integration.)
Even without taking the continuum limit,
the terms linear and quadratic in fluctuations 
have essentially the same structure to that of 
the commutative limit (\ref{topological}). 
One can also use the reflection symmetries 
(\ref{reflect1}), (\ref{reflect2})
to obtain the same conclusion.

\begin{figure}
\begin{center}
 \leavevmode
 \epsfxsize=70mm
 \epsfbox{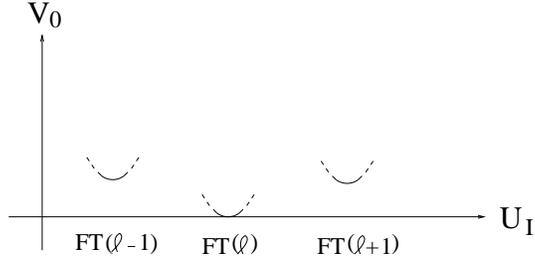}
\end{center}
\caption{Schematic figure 
of the potential $V_0 (U_I;\theta)$
for $\theta = 2\pi \ell /N$.
FT($\ell$) denotes the fuzzy torus configuration
(\ref{FTl}).}
\label{thetaell}
\end{figure}

\begin{figure}
\begin{center}
 \leavevmode
 \epsfxsize=70mm
 \epsfbox{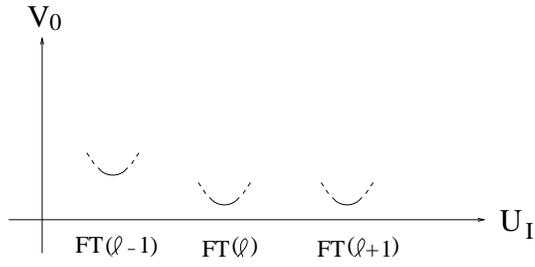}
\end{center}
\caption{The potential $V_0 (U_I;\theta)$
for $2\pi \ell /N < \theta < 2\pi (\ell+1) /N$.
Among the fuzzy torus minima,
the one whose 
$\theta_\ell$ is closest to $\theta$
has the smallest energy.}
\label{thetamiddle}
\end{figure}

\begin{figure}
\begin{center}
 \leavevmode
 \epsfxsize=70mm
 \epsfbox{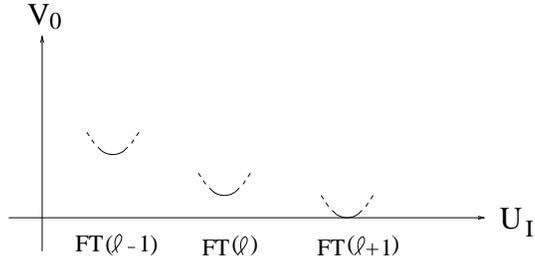}
\end{center}
\caption{The potential $V_0 (U_I;\theta)$
for $\theta = 2\pi (\ell+1) /N$.}
\label{thetaplus}
\end{figure}

The energy of the local minimum $U_I = V_I(\ell)$
is calculated to be
\bea
 \label{vacE}
V_0 (V_I;\theta)
=
- 2 g^2 f^4 |c_0| kN \cos \delta \theta .  
\eea
This means that
among the fuzzy torus minima labeled by $\ell$,
the one whose 
$\theta_\ell$ is closest to $\theta$
has the smallest energy.
See Figs.~\ref{thetaell}-\ref{thetaplus}.
Although we have not completely sought out 
the all minima of the potential (\ref{potth}), 
from (\ref{vacE})
it seems reasonable to assume that the fuzzy torus configuration
with $\theta_\ell$ closest to $\theta$ is the 
absolute minimum.
If this is the case, no fine tuning for $\theta$ is required.
Notice that when $N$ is large, there is always $\theta_\ell$
which is close  
($\delta \theta \lesssim 2\pi / N$)
to $\theta$.

The fuzzy torus looks closer to the ordinary torus
when $\ell / N$ is small.
By assuming that the CP violation (or the violation of the reflection symmetry) is small,
$\theta \ll 1$ is preferred,
thus small $\ell / N$ is preferred.


\subsubsection*{Comment on the possibility of dynamical tuning of $\theta$}

There is another scenario with an attractive feature,
which however has a problem in the naturalness.
We will explain this possibility below.

Let us consider the case where $\theta$ is not a constant but
a dynamical field depending on the four-dimensional coordinates $x$:
\bea
 \label{potthx}
V_0(U_I(x), \theta(x) ) = g^2 f_1^2 f_2^2 |\tc_0| \tV_0 (U_I(x),\theta(x)) ,
\eea
where
\bea
 \label{tpotthx}
\tV_0 (U_I(x),\theta(x))
\equiv 
-
\tr_{SU(kN)} 
\left[
e^{i \theta}
U_1 U_2 U_1^\dagger U_2^\dagger
+
e^{-i \theta}
U_2 U_1 U_2^\dagger U_1^\dagger
\right].
\eea
Then, the combination 
\bea
 \label{comb}
\theta = \frac{2\pi}{N} \, \ell, \quad 
U_I = V_I(\ell) \quad
(\ell: \mbox{integer}),
\eea
is the minimum of the potential (\ref{potthx}).
In string theory,
the non-commutative parameter 
usually arise from form field background,
and thus 
it is natural to expect that it becomes dynamical at some energy scale.
An attractive feature of this scenario is that
this provides a model for the
spontaneous CP symmetry breaking.

However, regarding (\ref{potthx}) as the leading
potential
is problematic from the point of view of effective field theory.
The following term can be induced by the quantum correction
\bea
 \label{qpot}
V_q(U_I) 
= 
c_q \tr 
\left[
U_1U_2U_1^\dagger U_2^\dagger
\right] + h.c.
\eea
where the natural magnitude of the complex coefficient $c_q$
is estimated to be
\bea
c_q \sim g^2 f^4.
\eea
The combined potential $V_0 + V_q$ no longer 
has the combination (\ref{comb}) as its minimum.

The problem is that no symmetry guarantees
the form of the 
potential (\ref{potthx}) to be the leading term
in our effective field theory.
The best one can do may be 
to promote the imaginary part of $\tc_0$ 
in (\ref{tpotth}) to be dynamical.
The real part of $\tc_0$ 
is a coefficient of the CP even term and
the imaginary part is that of the CP odd term.
Thus the separation of the real part and imaginary part
of $\tc_0$ is protected from the quantum corrections
by the CP symmetry,
and it is possible to promote only the imaginary part dynamical.
The quantum correction (\ref{qpot})
contributes to $\Im \, \tc_0$ as a constant shift.
This can be absorbed into the field redefinition
of $\Im \, \tc_0$ if one requires this shift to be a
symmetry of the other part of the action.
This is reminiscent of the
Peccei-Quinn symmetry for the axion \cite{Peccei:1977hh}.
However, 
here we are interested in 
the CP breaking vacua.
When $\theta \ll 1$,
$\theta \sim \Im \, \tc_0 / \Re \, \tc_0$
and 
in this case the combination
(\ref{comb})
may become an approximate solution,
with dynamical $\Im \, \tc_0$ helping for
lowering the potential energy.\footnote{%
While for the leading potential 
(\ref{potth}) with $\Im \, \tc_0$ dynamical
the configuration
$\Im \, \tc_0$ = 0 will be the minimum of the potential, 
it might be modified
by the higher order corrections.}

\subsection{Arbitrary loop diagrams
and the symmetry constraints}\label{secorder}

\subsubsection*{Analysis at small $N$}

We first consider the case 
when $N$ is small and 
can be neglected for a rough order estimate.
The $N$ dependence will be incorporated
after this analysis.

What is important for the suppression of
the quantum corrections
to the zero-modes 
is the
$SU_L(kN) \times SU_R(kN)$ 
subgroup of the
weakly broken chiral symmetry (\ref{chiral}).
The breaking of the approximate 
chiral symmetry 
can be parametrized by introducing
non-propagating
``spurion" $s$
which takes value in $kN \times kN$ complex matrix.
We assign transformation laws for $s$
under the chiral transformation
so that the covariant derivatives transform
homogeneously under the chiral transformation.
Thus we replace the covariant derivative 
(\ref{CD}) as follows:
\bea
D_\mu U_I 
=
\frac{1}{g} 
\pa_\mu U_I s 
- i  A_\mu U_I s + i U_I s A_\mu .
\eea
Then we can make the
$SU_L(kN) \times SU_R(kN)$ 
part of the $(SU_L(kN) \times SU_R(kN) )^2$
chiral transformation
to a global symmetry of the action:
\bea
\label{spurious}
U_I &\rightarrow& L U_I R^\dagger ,\\
A_\mu &\rightarrow& L A_\mu L^\dagger ,\\
s &\rightarrow&      R s L^\dagger    .
\eea
After using the spurion $s$ 
to define the global chiral symmetry,
we set $s = g \mathds{1}_{kN}$.
This breaks the chiral $SU_L(kN) \times SU_R(kN)$
to its diagonal subgroup in which $L=R$.

Now, notice that 
in the fuzzy torus vacuum (\ref{UV}),
the vacuum expectation value
of $\tr_{SU(kN)} \, U_I{}^\ell $
vanishes except for 
the special value of $\ell$
due to the remaining $\Z_N \times \Z_N$ symmetry
(\ref{Zn}) on the vacuum:
\bea
 \label{onfuzzT}
\tr_{SU(kN)} \, {U_1}^\ell
\Bigr|_{U_I=V_I}
= 0 \quad (\ell \ne (\mbox{multiple of } N) ).
\eea
Thus in the leading order
in the power series expansion in $1/ \Lambda$,
the operator which is consistent with 
the global symmetry
${\mathbb Z}_{kN} \times {\mathbb Z}_{kN}$ 
(\ref{ZNk})
and the spurious symmetry (\ref{spurious})
which contributes to the mass term of 
the zero mode $u_0^1(x)$
is
\bea
 \label{highert}
\left|
\tr_{SU(kN)} \,
\left(U_1 s\right)^N
\right|^2  .
\eea
The analysis for the mass of the zero-mode $u_0^2(x)$ is similar.
The operators 
$\left|
\tr_{SU(kN)} \,
\left(U_1 s\right)^\ell
\right|^2$ with $\ell \ne$ (multiple of $N$)
can be generated in the effective action,
but they do not contribute to the mass term
of the zero modes due to (\ref{onfuzzT}).
From the identification of the fields $U_I$
with the link variables in the lattice gauge theory,
the operator (\ref{highert}) can be regarded as  
the square of the Wilson loop wrapped on 
the 1-cycle of the torus with the radius $R_1$.

After setting $s = g \mathds{1}_{kN}$,
(\ref{highert})
is proportional to $g^{2N}$.
Comparing with (\ref{order}) in the appendix \ref{estimate},
the power factor $2N$ on $g$ 
should be equal to the power $2I_A - 2G$ on the
suppression factor $f/\Lambda$, 
where 
$G$ is the number of purely gauge interaction vertices and 
$I_A$ is the number of the gauge field propagators in
a Feynman diagram under consideration.
Thus the contribution to
the coefficient of the operator (\ref{highert})
from the $L$-loop Feynman diagrams
is estimated as
\bea
 \label{coe}
 f^2 \Lambda^2
\left(
\frac{\Lambda}{4\pi f}
\right)^{2L}  \left( \frac{gf}{\Lambda} \right)^{2N},
\eea
with a dimensionless numerical coefficient of
order one.

From (\ref{coe}),
we observe that 
at the one loop $L=1$,
the quadratic dependence on the cut-off $\Lambda$
only appears
when $N = 1$,
and the logarithmic dependence appears
when $N =2$.
However, 
for $N=1$ we do not have the fuzzy torus solution. 
Therefore
there is no quadratic divergence 
in the radiative corrections to the mass of the zero-modes.
On the other hand, 
our one-loop results
(\ref{oneloop}) and (\ref{divstr})
are consistent with the above argument:
When $N=2$ there is a log divergence
while for $N \geq 3$ 
the radiative corrections to 
the mass of the zero-modes is finite.

Before setting $\Lambda = 4\pi f$
(since we are considering small $N$ case here
we neglect the factor $1/\sqrt{N}$)
the expression (\ref{coe}) is seemingly more divergent
for higher loop contributions,
since it is proportional to $\Lambda^{2+2L-2I_A}$.
However, our effective field theory
is valid up to the cut-off scale $\Lambda \approx 4\pi f$.
At this cut-off scale all loops
contributes in the same order in the magnitude.
After setting $\Lambda = 4\pi f$,
the coefficient (\ref{coe}) is estimated as
\bea
f^4
16 \pi^2
\left(
\frac{g}{4\pi}
\right)^{2N} .
\eea
This is the natural magnitude of the coefficient for 
the operator (\ref{highert}).

To summarize, two global symmetries played
the major roles in suppressing 
the quantum corrections to the 
mass of the zero modes:
the weakly broken 
chiral symmetry
introduces the suppression factor 
${gf}/{\Lambda}$,
while how much powers are on the suppression factor
is determined
by the unbroken $\Z_{N} \times \Z_{N}$ symmetry (\ref{Zn}).

The natural magnitude of the coefficient 
$c_0$
in the action (\ref{action}) 
can be estimated in a similar way
to give the explained order.

\subsubsection*{%
Incorporating $N$ dependence}\label{seccont}

When comparing the fuzzy torus with different $N$,
we should fix the radii of the fuzzy torus:
\bea
 \label{fuzzlatt}
2\pi R_I = a_I N : fixed \quad (I =1,2).
\eea
This means when we take $N$ to be large,
we should scale $a_I$ as
\bea
 \label{ascale}
a_I \sim \frac{1}{N} .
\eea 
$1/ a_I = N /R_I$ is the energy scale
where the discrepancy between the fuzzy torus
and the ordinary torus becomes large.
Thus ``divergences" associated with the limit $N \rightarrow \infty$
is related to the
UV divergences in the extra dimensional directions.\footnote{%
Here we use the term ``divergence" in the same sense to that in the footnote
\ref{footdiv}.
While we will not take $N$ to infinity,
the dependence on the large $N$ 
is the dependence on the UV scale in the extra dimensional directions.}
We should also compare the theory
with the same four-dimensional
$SU(k)$ gauge coupling.
Thus 
we obtain the scaling\footnote{%
It is interesting to observe that this scaling
is the same to that of the (de)construction with 
one-dimensional periodic lattice rather than that of the
two-dimensional periodic lattice.}
\bea
 \label{gscale}
\frac{g}{\sqrt{N}} \equiv g_{SU(k)} : fixed.
\eea
%
Together with (\ref{ascale}) and (\ref{agf}), this means
\bea
 \label{fscale}
\frac{f_I}{\sqrt{N}} : fixed .
\eea

Taking into account the $N$ dependence
as (\ref{loopN}) in the appendix \ref{estimate},
the coefficient (\ref{coe}) 
of the operator (\ref{highert})
which leads to the mass of the zero-modes
is modified as 
\bea
 \label{coeN}
\frac{ f^2 \Lambda^2 }{kN}
\left(
\frac{\Lambda \sqrt{kN}}{4\pi f}
\right)^{2L}  \left( \frac{gf}{\Lambda} \right)^{2N}.
\eea
The first $1/kN$ factor comes from the fact
that (\ref{highert}) is a
double trace operator.

We restrict ourselves to the case where
the factor $gf/\Lambda$ is small and 
provides a suppression factor in (\ref{coeN}),
which in turn suppresses the mass of the zero-modes.
From the definition (\ref{agf}),
it amounts to the case
when the highest end of the mass spectrum around the fuzzy torus vacuum
is 
below
the UV cut-off scale $\Lambda$ determined from 
the validity of the loop expansion:
\bea
 \label{lmdub}
\Lambda = \frac{4 \pi f}{\sqrt{kN}} 
\gg
gf
=
\frac{1}{a} =
\frac{N}{2\pi R} .
\eea
This provides the upper limit $N_{c}$ of $N$ 
around which 
the 
expansion in terms of 
$gf/\Lambda$
breaks down:
\bea
 \label{Nupbd}
N \ll N_{c} \approx \frac{4\pi}{g_{SU(k)} \sqrt{k}} .
\eea
In the application to the electroweak symmetry breaking,
the subgroup of $SU(k)$ should be identified
with the standard model gauge group $SU(2)\times U(1)$.
This would constrain the value of $g_{SU(k)} $ to be 
around $\sim 0.5$ 
(a typical value for an order estimate).
Thus we obtain a rough estimate for the upper limit $N_{c}$: 
\bea
 \label{Nc}
N_{c} \approx \frac{25}{\sqrt{k}} .
\eea

\section{A D-brane inspired model}\label{secD}

While UV completions are not necessary  
from the effective field theory point of view, 
they do provide good motivations for
the assumed symmetries in the effective field theory. 
As in the case of (de)construction \cite{ArkaniHamed:2001ca},
quiver gauge theories are possible UV completions
of the unitary matrix model discussed in the previous section.
Here we present another UV completion inspired by
the worldvolume theory on
D-branes probing ${\mathbb C}^3 /(\Z_N \times \Z_N)$ orbifold 
with discrete torsion.
A difference between the two UV completions
is that
chiral symmetry breaking of the quiver gauge theory
which is expected to lead to the unitary matrix model
via the non-linear realization
occurs at the strong coupling regime,
whereas 
the D-brane inspired model is perturbative.
%
Since
we are motivated by the fact that
fuzzy spaces are ubiquitously realized by D-branes,
the D-brane inspired model
is a natural direction to investigate.
Additionally,
D-branes
on ${\mathbb C}^3 /(\Z_N \times \Z_N)$
with discrete torsion
\cite{Ho:1998xh,Douglas:1998xa,Douglas:1999hq}
has an explanation
for the special value of $\theta$ (\ref{theta})
which is required for the existence of the
fuzzy torus vacuum.

Let us consider the following action:
\bea
\label{Zaction}
S
&=&
\int d^4 x\,
\tr_{SU(N\times k)} 
\Biggl[
-\frac{1}{2} F_{\mu\nu}(x) F^{\mu\nu} (x) \nn \\
&&\qquad 
+
\sum_{I=1}^3
\biggl\{
D_{\mu} Z_I (x) D^{\mu} Z_I^\dagger (x)
-
\frac{g^2}{2}
[Z_I, Z_I^{\dagger}]^2 
\biggr\}
\nn \\
&&\qquad 
-
g^2
\sum_{I = 1}^{3\, (mod\, 3)}
\Bigl[
e^{i \theta} Z_I Z_{I+1} - Z_{I+1} Z_I 
\Bigr]
\Bigl[e^{-i\theta} 
Z_{I+1}^{\dagger} Z_I^{\dagger} 
-Z_I^{\dagger} Z_{I+1}^{\dagger}
\Bigr]
\nn\\
&&\qquad 
-
\sum_{I=1,2}
\frac{M_I^2}{4 f_I^2} 
\left(
Z_I Z_I^{\dagger} 
-
f_I^2
\right)^2
- M_3^2 Z_3 Z_3^\dagger
\Biggr].
\eea
Here,
\bea
Z_1 = X_1 + i X_2 ,\quad
Z_2 = X_3 + i X_4 ,\quad
Z_3 = X_5 + i X_6,
\eea
where $X_I$ are
$kN \times kN$ Hermite matrices
which are adjoint representations of $SU(kN)$.
The covariant derivative for $Z_I$ is given by
\bea
D_\mu Z_I =
\pa_\mu Z_I - i g [A_\mu, Z_I] .
\eea
Except for the last line,
the action (\ref{Zaction}) is the bosonic part
of the low-energy effective action
realized on D-branes on an orbifold
${\mathbb C}^3 / (\mathbb{Z}_N \times \mathbb{Z}_N)$
with discrete torsion
\cite{Ho:1998xh,Douglas:1998xa,Douglas:1999hq}.\footnote{%
Similar action has been studied
as an extension of the
(de)construction to fuzzy spaces
\cite{Adams:2001ne,Dorey:2004iq}.}
At the tree level, $\theta$ takes the following discrete value:
\bea
\theta = \frac{2 \pi}{N} \, \ell.
\eea
This discreteness of $\theta$
at the tree level
is understood as 
discrete torsion of the 
$\C^3 / ( \mathbb{Z}_N \times \mathbb{Z}_N )$
orbifold.
As before, we will consider the case $\ell =1$ as an example.
%
%

The last line in the action (\ref{Zaction})
was introduced to
stabilize the radius of the 
fuzzy torus vacuum at the tree level,
so that scaler fields lighter than the model Higgs do not appear.

The action (\ref{Zaction}) has the following
global $U(1)^3$ symmetry:
\bea
 \label{rot}
Z_I \rightarrow e^{i \alpha_I} Z_I, \quad
(\alpha_I \in {\mathbb R}\mod 2\pi) .
\eea
This symmetry plays the similar role
to that of the $\Z_{kN} \times \Z_{kN}$ symmetry
in the unitary matrix model. 

The minimum of the 
potential is given by
\bea
\label{vacZ}
Z_1 = f_1 V_1,
\quad 
Z_2 = f_2 V_2 ,
\quad
Z_3 = 0 .
\eea
where the matrices $V_I$ are the same as the ones given in
(\ref{UV}).
The vacuum breaks the global $U(1)^3$ symmetry (\ref{rot})
to $\Z_N \times \Z_N \times U(1)$, as described around
(\ref{Zngauge}).

Now, any complex matrix can be 
decomposed by a unitary matrix and an Hermite matrix.
Thus we may decompose the fields $Z_I$ by 
unitary matrix $U_I$ and an Hermite matrix $H_I$ as follows:
\bea
 \label{polard}
Z_I = U_I H_I \quad(\mbox{for } I =1,2).
\eea
This decomposition is convenient and thus appropriate
for perturbative analysis
around the vacuum (\ref{vacZ}).
The unitary matrices $U_I$ plays the similar role
to the unitary matrix fields $U_I$ in the previous section
and thus we have used the same symbols.
The covariant derivative satisfies the Leibniz rule:
\bea
D_\mu Z_I =
(D_\mu U_I ) H_I + U_I (D_\mu H_I)
\quad(\mbox{for } I =1,2),
\eea
where 
\bea
D_\mu U_I &=& \pa_\mu U_I - i g [A_\mu, U_I], \\
D_\mu H_I &=& \pa_\mu H_I - i g [A_\mu, H_I] .
\eea
We can
extend the weakly broken global 
$SU_L(kN) \times SU_R (kN)$ chiral symmetry
on the unitary matrix model
discussed in section \ref{secorder}
to the current model:
\bea
&&U_I \rightarrow L U_I R^\dagger , \quad 
H_I \rightarrow R H_I R^\dagger \quad(\mbox{for } I =1,2),\nn \\
&&Z_3 \rightarrow L Z_3 R^\dagger, \nn\\
&&A_\mu \rightarrow L A_\mu L^\dagger, \quad s \rightarrow R s L^\dagger  .
\eea
When using the spurion $s$ to describe 
the chiral symmetry, the covariant derivatives 
for $H_I$ and $Z_3$ are modified to
\bea
D_\mu H_I &=&  \pa_\mu H_I + 
\frac{1}{g} \left( i s^\dagger A_\mu s H_I - i  H_I s^\dagger A_\mu s \right) ,\\
D_\mu Z_3 &=& \frac{1}{g} \pa_\mu Z_3 s - i A_\mu Z_3 s + i Z_3 s A_\mu . 
\eea
Since this model has the same symmetries to 
those in the unitary matrix model,
the mass of the zero-modes is suppressed
by essentially the same mechanism.

Notice that when $M_I \ll \Lambda$,
this model is not just a UV completion but
also introduces other fields to the unitary matrix model.
In order for the additional fields not to be lighter or have similar mass
to the zero-modes, we may require
\bea
M_I \gtrsim \frac{1}{R} .
\eea
On the other hand, we would
also like to require that 
the coupling constant
in front of the quartic coupling
$\tr (Z_IZ_I^\dagger)^2$
in the last line of (\ref{Zaction})
remains in the perturbative regime.
Thus we require
\bea
 \label{Zpert}
M_I \lesssim f_I . 
\eea
One may also like to require that the modification
from this model appears 
before the perturbative expansion of the
unitary matrix model breaks down.
This gives the bound
\bea
 \label{Upert}
M_I \lesssim \Lambda .
\eea
However, (\ref{Zpert}) tends to give stronger constraint,
as we have seen in section \ref{secorder}.

This model without the last line in (\ref{Zaction})
is an asymptotically free gauge theory
and theoretically
we can take its UV cut-off to infinity.
As a solution to the
Higgs naturalness problem,
this is an advantageous feature. 
However, the model has its own naturalness problem
due to the last line of (\ref{Zaction}),
since the masses $M_I$ of the scalar fields $Z_I$ are not protected by any symmetry.
This point may be refined by 
considering a different stabilization mechanism
for the fields $H_I$ in (\ref{polard}). 
Since this is a model dependent detail,
we leave this issue to the future investigations.
We expect the idea of the
gauge-Higgs unification in spontaneously created
fuzzy extra dimensions
to be general and have rich varieties of realizations.

\section{Discussions}\label{secdiss}


In this paper we focused on the naturalness issue 
regarding the mass of the Higgs field.
We hope our results provide a basis 
for the construction of more realistic models of 
the electroweak symmetry breaking.
In order to construct more realistic models, 
one should introduce the standard model fermions.
Notice that as in the ordinary gauge-Higgs unification,
the coupling between the fermions and the Higgs field
are tightly constrained by the gauge symmetry.
Fuzzy spaces are known to give additional constraints
to the possible gauge group and 
the representations of the matter fields.\footnote{%
See e.g. \cite{Aoki:2008ik} for more about the issue and 
some direction in the case of the fuzzy torus.}
It will be interesting to examine
how much of the mechanisms
employed in the gauge-Higgs unification
in ordinary extra dimensions
can be extended to the case of the
fuzzy extra dimensions.
We would like to point out that
the fuzzy extra dimensions may give rise to
interesting Yukawa texture, as has been discussed in
\cite{Cecotti:2009zf,Marchesano:2009rz,Heckman:2010pv,Furuuchi:2010gu}.
Moreover, the gauge field in the extra dimensions
is one of the candidates for the Higgs field in these models.
So far the weak-scale supersymmetry has been employed
to solve the Higgs naturalness problem in these models.
Our work may provide an economical alternative 
solution to the Higgs naturalness problem in such scenarios,
because the fuzzy extra dimensions are already 
built in in these scenarios.

In this work we studied the models with
the fuzzy torus extra dimensions.
The fuzzy torus might be special 
in that it circumvent the
following issue.
In the D-brane setting, 
the fuzzy space appears as a
vacuum solution for 
the matrix version of 
the embedding coordinate fields of D-branes.
In the situation where the
fuzzy space is embedded in higher dimensional space,
one would need to separate the fluctuations of the
matrix coordinate fields around 
the fuzzy space vacuum
into the gauge field components and the scalar field components.
This is because in the gauge-Higgs unification,
the zero-modes of the gauge field in the extra dimensions
are to be identified with the Higgs field.
The above separation amounts to the separation of the matrix coordinate fields
into the direction tangent to the fuzzy space and
perpendicular to the fuzzy space.
However, it is not clear how to make such separation
when the fuzzy space is made from finite size matrices.
On the fuzzy space described by finite size matrices,
the KK modes on the fuzzy space are truncated at some finite level.
This means that we may not have enough functions 
to make a coordinate transformation
which is needed for the separation
of the matrix coordinate fields in higher dimensions into
the components tangent- and perpendicular- 
to the embedded fuzzy space.
Moreover, we have to make the coordinate transformation
with the non-commutative 
matrix product.\footnote{%
See \cite{deBoer:2003cp,Brecher:2004qi} and references
for a closely related issue.}
Let us explain with the fuzzy sphere \cite{Madore:1991bw}
as an example.
\begin{figure}
\begin{center}
 \leavevmode
 \epsfxsize=70mm
 \epsfbox{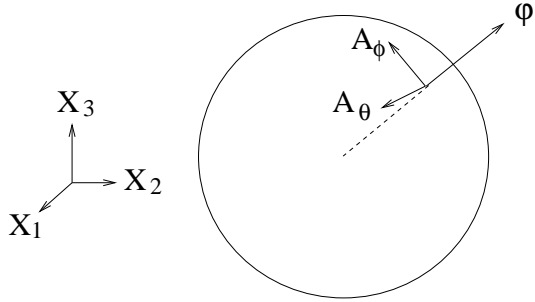}
\end{center}
\caption{Decomposition of the vector in $\mathbb{R}^3$
(parametrized by the Cartesian coordinates $X^1,X^2,X^3$)
into
components tangent to the sphere ($A_\theta,A_\phi$)
and perpendicular to the sphere ($\varphi$).
The sphere is embedded in the Euclidean space $\mathbb{R}^3$.
It is not clear how to make such decomposition
for the fuzzy sphere with finite size matrix coordinates.}
\label{sphere}
\end{figure}
Fig. \ref{sphere} is a figure of a sphere
embedded in $\mathbb{R}^3$.
The matrix coordinate fields $X^I$ ($I=1,2,3$) of D-branes
are associated with
the Cartesian coordinates 
of $\mathbb{R}^3$ in the figure.
The fuzzy sphere is described by the matrix coordinate fields
satisfying
\bea
 \label{fsphere}
[X^I,X^J] = i \alpha \epsilon^{IJK} X^K ,
\eea
where matrix coordinate fields $X^I$
are Hermite matrices and $\alpha$ is a real number.
To discuss gauge-Higgs unification 
in the fuzzy sphere extra dimensions,
one should extract the zero-modes of 
the gauge fields 
from the fluctuations around the (\ref{fsphere}).
However, as explained above,
it is not clear how to extract 
the gauge field on the fuzzy sphere
when the size of the matrices are finite 
\cite{CarowWatamura:1998jn,Iso:2001mg,Steinacker:2003sd}.
The fuzzy torus was little special
in that it can be described by unitary matrices,
and any complex matrix can be decomposed
into a product of a unitary matrix and an Hermite matrix,
see (\ref{polard}).
Thus the generalizations of the gauge-Higgs unification
to other fuzzy extra dimensions
remain as interesting future directions. 
On the other hand, torus has been one of the most 
useful backgrounds in the ordinary gauge-Higgs unification,
and the fuzzy torus may also remain as the
most basic background in the study of 
the gauge-Higgs unification in fuzzy extra dimensions.
One possibility may be that
the gauge-Higgs unification picture may not be necessary
and one may do without the separation mentioned above.
When $N$ is small the extra dimensions 
look far from
the ordinary space.  
And in the general discussions in section \ref{secorder}, 
we were mostly working with the Wilson lines 
rather than the gauge field on the fuzzy space,
and we may generalize pushing along this direction.


\vskip7mm
\noindent
\centerline{\bf Acknowledgments}\vskip1mm
The authors thank
Pei-Ming~Ho,
Hirotaka~Irie,
Yoshiharu~Kawamura,
C.~S.~Lim,
Chia-Min~Lin,
Eibun~Senaha,
Tomohisa~Takimi, 
Dan~Tomino and
Satoshi~Watamura
for useful discussions.
K.F. thanks IPMU for the support and the hospitality during his visit.
He also thanks Particle Physics Group of Chuo University for the hospitality.
T.I. benefitted much from his visits 
to NCTS (Hsinchu), NCTS (North) and NTU CTS 
in the course of finalizing this work.
T.I. is partially supported by the grants of the Ministry 
of Education, Kiban A, 21540278 and Kiban C, 21244063.
K.O. would like to thank Hajime Aoki, Naoki Sasakura
and Asato Tsuchiya for valuable discussions during
the YITP workshop ``Field Theory and String Theory".
K.O. is supported in part by JSPS Grant-in-Aid
for Young Scientists (B) 23740178.

\appendix

\section{Fuzzy torus and emergent gauge field}\label{AfuzzT}

\subsection{Fuzzy torus}

Fuzzy torus is
described by unitary matrices
$W_1$, $W_2$
subject to the relation
\bea
W_1 W_2 = e^{-i \theta} W_2 W_1 ,
\eea
where
\bea
 \label{Ashift}
\theta = \frac{2\pi}{N} .
\eea
Here, $N$ is the size of the matrices $W_1$ and $W_2$,
which is a parameter of the fuzzy torus.
An explicit realization
of $W_1$ and $W_2$ satisfying (\ref{Ashift}) is given by
the so-called 't Hooft-Weyl matrices:
\bea
W_1 = 
\left(
\begin{array}{ccccc}
1 & & & & \\
 & e^{- i\theta} & & & \\
 & & e^{- i 2 \theta}  & & \\
 & & & \ddots & \\
 & & & & e^{- i (N-1) \theta}
 \end{array}
\right),
\quad
W_2 = 
\left(
\begin{array}{ccccc}
0 & 1 & & & \\
 & 0 &1 & & \\
 & & \ddots & \ddots &  \\
 & & & 0 & 1 \\
1 & & &  & 0 
 \end{array}
\right) ,
\eea
(the empty entries should be read as zero).
Any $N \times N$ matrix $\varphi$
can be expanded in terms of $W_1$ and $W_2$,
which can be interpreted as the
Fourier expansion on the fuzzy torus:
\bea
 \label{FF}
\varphi = \sum_{m} \sum_{n}
\varphi_{(m,n)} e^{i mn \theta} W_1^m W_2^n 
\times \frac{1}{\sqrt{2N}},
\eea
Here,	$T^{(m,n)} \equiv e^{i mn \theta} W_1^m W_2^n 
\times \frac{1}{\sqrt{2N}}$
are normalized so that
$\tr \, T^{(m,n)} T^{(m',n')} = \frac{1}{2} \delta_{m+m',0} \delta_{n+n',0}$.
$m$ and $n$ in the summation in (\ref{FF})
run over integers in $ -\frac{N}{2} \leq m,n < \frac{N}{2}$.
The phase factor in (\ref{FF}) is chosen so that
for an Hermite matrix $\varphi = \varphi^\dagger$,
$\varphi_{(m,n)} = \varphi_{(-m,-n)}^\ast$.
The inverse transformation of (\ref{FF})
is given by
\bea
 \label{invFF}
\varphi_{(m,n)} = 
\sqrt{\frac{2}{N}} \tr 
\, \varphi \, W_1^{-m} W_2^{-n} .
\eea
As can be foreseen from
calling it as 
Fourier expansion on the fuzzy torus,
$W_{1}$ and $W_2$ are analogous to 
$e^{- i \frac{\phi_1}{R_1}}$ and
$e^{  i \frac{\phi_2}{R_2}}$ respectively,
where
$\phi_1$ and $\phi_2$ are the periodic coordinates
on ordinary torus:
$\phi_{1} \sim \phi_{1} +  2 \pi R_{1}$,
$\phi_{2} \sim \phi_{2} +  2 \pi R_{2}$.

From (\ref{Ashift}), it follows that
\bea
 \label{diff}
W_1 (W_1^m W_2^n) W_1^{\dagger} &=& e^{-i n \theta} (W_1^m W_2^n) , \nn \\
W_2 (W_1^m W_2^n) W_2^{\dagger} &=& e^{i m \theta} (W_1^m W_2^n).
\eea
This means that 
the unitary transformation by $W_1$ generates
an analogue of the translation of the $\phi_2$ coordinate
by $- \theta R_2$, 
and the unitary transformation by $W_2$
generates
an analogue of the translation by $\theta R_1$ for the $\phi_1$ coordinate.

Thus if we define the difference operators
\bea
 \label{fdiff}
\delta_1 \varphi 
&\equiv&
W_1 \varphi W_1^\dagger - \varphi , \nn \\
\delta_2 \varphi 
&\equiv&
W_2 \varphi W_2^\dagger - \varphi ,
\eea
it follows from (\ref{diff}) that
\bea
 \label{delta}
\delta_1 (W_1^m W_2^n) 
&=& 
(e^{-i n \theta} - 1)(W_1^m W_2^n)
= - 2 i e^{-\frac{i n\theta}{2}} \sin \left( \frac{n \theta}{2} \right) 
(W_1^m W_2^n) ,
\nn \\
\delta_2 (W_1^m W_2^n) 
&=& 
(e^{i m \theta} - 1)(W_1^m W_2^n)
= 2 i e^{\frac{i m\theta}{2}}  \sin \left( \frac{m \theta}{2} \right) 
(W_1^m W_2^n).
\eea
In the commutative limit $N \rightarrow \infty$,
$W_1$ and $W_2$ can be identified
with the commutative periodic 
coordinates
$\phi_1$, $\phi_2$ on the torus 
($\phi_{1,2} \sim \phi_{1,2} +  2 \pi R_{1,2}$) as
\bea
 \label{identify}
W_1 \rightarrow e^{- i \frac{\phi_1}{R_1}}, \quad
W_2 \rightarrow e^{  i \frac{\phi_2}{R_2}} .
\eea
This is based on 
the following algebraic relation in
the $N \rightarrow \infty$ limit
with fixed fuzzy torus radii: 
\bea
 \label{aftr}
2\pi R_I = N a_I  : fixed,
\eea
\bea
 \label{fuzzderi}
\frac{1}{a_1} \delta_1 W_2^n
&\rightarrow &
-  i \frac{n}{R_2} W_2^n
\leftrightarrow \pa_{\phi_2} e^{- i \frac{n}{R_2} \phi_2},\nn \\
\frac{1}{a_2}
\delta_2 W_1^m
&\rightarrow &
i \frac{m}{R_1} W_1^m
\leftrightarrow
\pa_{\phi_1} e^{ i \frac{m}{R_1} \phi_1} .
\eea
where $\leftrightarrow$ indicates that
the same algebraic relations
are satisfied with the identification
(\ref{identify}).
Thus in the $N \rightarrow \infty$ limit,
those algebraic relations reduce to
those of differentiations on 
the periodic functions 
on a commutative torus.
On the other hand,
the trace becomes the integration on the torus: 
\bea
 \label{FTint}
\frac{1}{N} \tr \rightarrow 
\int 
\frac{d\phi_1}{2\pi R_1} 
\frac{d\phi_2}{2\pi R_2} .
\eea

\subsection{Emergent gauge field on the fuzzy torus}

In the ordinary gauge-Higgs unification,
the inhomogeneous part of the local gauge transformation
forbids the mass term of the gauge field.
Therefore, 
it would be useful to observe that
the fluctuations around the fuzzy torus vacuum 
contain an excitation which can be identified with 
the components of the gauge field
in the fuzzy torus directions.

\subsubsection*{Emergence of the inhomogeneous gauge transformation}

Let us expand the fields $U_I$ around the 
fuzzy torus background (\ref{UV}):
\bea
 \label{Uexp}
U_I = e^{i a_I \A_I} V_I  .
\eea
Originally, the fields $U_I$ transform homogeneously under
the $SU(kN)$ local gauge transformation:
\bea
U_I \rightarrow e^{i \lambda} U_I e^{- i \lambda},
\eea
where $kN \times kN$ Hermite matrix
$\lambda(x)$ is a gauge transformation parameter.
The inhomogeneous gauge transformation
of the 
gauge field components in the fuzzy torus directions
appear by requiring that 
the form (\ref{Uexp}), i.e.
the separation of the fuzzy torus background part
and the fluctuation part is fixed under the
$SU(kN)$ gauge transformation.
\bea
U_I = e^{i a_I \A_I} V_I 
\rightarrow
e^{i \lambda} 
\left(
 e^{i a_I \A_I}  V_I
\right) 
e^{- i \lambda}
=
e^{i \lambda}  e^{i a_I \A_I} e^{- i \lambda'} V_I ,
\eea
where
\bea
e^{-i \lambda'} 
\equiv 
V_I e^{-i \lambda} V_I^{-1}  = e^{-i V_I\lambda V_I^{-1}},
\eea
or
\bea
\lambda' = V_I \lambda V_I^{-1} .
\eea
By the requirement that the
form of our parametrization (\ref{Uexp}) is fixed,
the gauge transformation law for the fluctuation
$\A$ $\rightarrow$ $\A^{\lambda}$ should be defined by
\bea
e^{i a_I  \A_I^{\lambda}} =  e^{i \lambda}  e^{i a_I \A_I} e^{- i \lambda'}.
\eea
For an infinitesimal $\lambda$,
\bea
e^{i a_I \A^\lambda} 
&=&  e^{i \lambda}  e^{i a_I \A_I} e^{- i \lambda'} \nn \\
&=&  e^{i \lambda}  e^{i a_I \A_I} e^{-i \lambda} e^{i \lambda} e^{- i \lambda'} \nn\\
&=& \exp \left[  i a_I \A_I - i V_I \lambda V_I^{-1} + i \lambda 
    - a_I [\lambda, \A_I ] + {\cal O} (\lambda^2) \right] \nn\\
&=& \exp \left[ i a_I \A_I - i \delta_I \lambda - a_I [\lambda, A_I] 
+ {\cal O} (\lambda^2) \right],
\eea
where $\delta_I$ is defined in (\ref{fdiff}).
Thus in terms of the field $\A_I$,
the gauge transformation is given by
\bea
 \label{inhom}
\A_I 
\rightarrow 
\A_I^\lambda =
\A_I - \frac{1}{a_I}\delta_I \lambda + i [\lambda, A_I] + {\cal O} (\lambda^2).
\eea
Recalling that $\frac{1}{a_I}\delta_I$
can be regarded as a derivative on the fuzzy torus
(\ref{fuzzderi}),
we can regard (\ref{inhom}) as
the inhomogeneous $SU(k)$ gauge transformation
for the gauge field $\A_I$ on the fuzzy torus.

\subsubsection*{Wilson loop operator}

Next we explain
why the operator
\bea
 \label{UN}
\frac{1}{kN}\,
\tr \, {U_I}^N  ,
\eea
can be regarded as the
Wilson loop operator on the fuzzy torus \cite{Ambjorn:1999ts}.
Using (\ref{Uexp}), (\ref{UN}) can be rewritten as
\bea
 \label{fdwilson}
&& kN \, \tr_{SU(kN)} \, {U_I}^N \nn\\
&=&
\tr_{SU(kN)} 
\left[ \left( V_I e^{i a_I \A_I} \right)^N 
\right] \nn\\
&=&
\tr_{SU(kN)}
\left[
\left(V_I e^{i a_I \A_I} V_I^{-1} \right)
\left(V_I^2 e^{i a_I \A_I} V_I^{-2} \right)
\cdots
\left(V_I^N e^{i a_I \A_I} V_I^{-N} \right) 
\right]\nn\\
&=&
\tr_{SU(kN)} \left[
 e^{i a_I \A_I}
\left(V_I e^{i a_I \A_I} V_I^{-1} \right)
\left(V_I^2 e^{i a_I \A_I} V_I^{-2} \right)
\cdots
\left(V_I^{N-1} e^{i a_I \A_I} V_I^{-(N-1)} \right) 
\right],\nn\\
&&
\eea
where we have used
\bea
V_I^N = 1 ,
\eea
and the cyclic property of the trace.
From the correspondence between the 
fuzzy torus and the ordinary torus
through the $N \rightarrow \infty$ limit
with $a_I N = 2\pi R_I$ fixed discussed previously,
(\ref{fdwilson}) can be
regarded as a discretized version of 
\bea
\lim_{N \rightarrow \infty}
\tr_{SU(kN)} \left[
    e^{i a_I \A_I (\phi_I)}
    e^{i a_I \A_I (\phi_I + a_I)}
    e^{i a_I \A_I (\phi_I + 2 a_I)}
     \cdots
    e^{i a_I \A_I (\phi_I + (N-1) a_I)}
    \right] .
\eea
Rewriting using (\ref{FTint}),
this formally has the form of the Wilson loop operator
wrapping the $I$-th direction once and
integrated over the fuzzy torus:
\bea
\int \frac{d\phi_1}{2\pi R_1} \frac{d\phi_2}{2\pi R_2}
\frac{1}{k}
\tr_{SU(k)}P \exp i \oint \A_I \phi_I ,
\eea
Here, $\tr_{SU(k)} P$ denotes the path ordered trace.
However, $N \rightarrow \infty$ limit is a little bit formal 
since for any finite $N$, one needs
the integration over the fuzzy torus
corresponding to the trace of $U(N)$ gauge subgroup
in order for the operator to be gauge invariant observable.

\section{The estimate of the UV cut-off scale $\Lambda$}\label{estimate}


In this appendix,
we estimate
the natural UV cut-off scale $\Lambda$ for the
effective field theory (\ref{action}).
We will follow the argument of \cite{Manohar:1983md}.

Let us parametrize the unitary matrix fields as
\bea
U_I =  e^{i \frac{\pi_I}{f}} V_I,
\eea
where $V_I$ is the vacuum expectation value of $U_I$.\footnote{%
The difference from the parametrization in (\ref{Uexp})
is just a matter of taste. 
In the appendix \ref{AfuzzT},
the analogy with the lattice gauge theory was useful
and therefore we used the notation 
closer to those often used in the lattice gauge theory. 
In this appendix, the analogy with the chiral perturbation theory
is useful, thus we use the notation similar to those used
in the description of pion.}
Consider general possible vertex
involving 
the gauge field $A_\mu$ and $\pi_I$.

\subsubsection*{Analysis at small $N$}

We first study the case when
$N$ is small 
and one can neglect the $N$ dependence
in the rough order estimate.
The $N$ dependence will be included 
after this analysis.

A coefficient consistent with naive dimensional analysis is
\bea
 \label{vertex}
(2 \pi)^4 \delta^4\left({\sum p} \right)
\left(  \frac{g A_\mu}{\Lambda} \right)^A
\left( \frac{\pi}{f} \right)^B
\left( \frac{p}{\Lambda} \right)^C
f^2 \Lambda^2 .
\eea
This correctly estimates
the coefficient
of the kinetic term for $U_I$
but underestimates the 
gauge field kinetic term.
This 
can be corrected
by multiplying the factor
\bea
\left(
\frac{\Lambda}{gf} 
\right)^2 ,
\eea
for the terms purely made from the gauge fields
and their derivatives.

Now, consider arbitrary Feynman diagram
involving a total of $V$
vertices of the form (\ref{vertex})
with $(A,B,C)$ values equals to 
$(A_i,B_i,C_i)$, $i=1,\cdots, V$.

The diagram simply gives
\bea
 \label{vertexc}
&&(2 \pi)^4 \delta^4\left({\sum p} \right)
\left(  \frac{g A_\mu}{\Lambda}  \right)^A
\left( \frac{\pi}{f} \right)^B
\left( \frac{p}{\Lambda} \right)^C
f^2 \Lambda^2
\nn \\
&\times&
g^{-A + \sum_i A_i} f^{B - \sum_i B_i + 2V-2}
\Lambda^{A + C + 2V-2 - \sum_i (A_i +C_i) } \nn \\
&\times&
k^{-C +\sum_i C_i }
(2\pi)^{4(V-1)}
\left[ \delta^{(4)} \left(\sum p_i \right) \right]^{V-1} 
\left[ \int \frac{d^4k}{(2\pi)^4} \frac{1}{k^2} \right]^I 
\left( \frac{\Lambda^2}{g^2 f^2} \right)^{G}  ,
\eea
where 
$G$ is the number of the purely gauge interaction vertices 
and $I$ is the number of 
the internal propagators
in the Feynman diagram.
Thus we use
\bea
&&\sum A_i = A + 2 I_A , \\
&&\sum B_i = B + 2 I_\pi ,
\eea
(conservation of ends of propagators)
where $I_A$ and $I_\pi$
are the number of internal propagators
of  $A_\mu$ and $\pi_I$
in the Feynman diagram, respectively,
and $I_A + I_\pi = I$.
We also have the equality
\bea
L 
  = I - V + 1  ,
\eea
where $L$ is the number of the loops
in the Feynman diagram.

Since all the momentum integrals
are cut off at $\Lambda$,
we can estimate them
by replacing all internal momenta by $\Lambda$:
\bea
k &\rightarrow & \Lambda ,\\
\left[ \int \frac{d^4k}{(2\pi)^4} \right]^L 
&\rightarrow &
\frac{\Lambda^{4L}}{(4\pi)^{2L}},
\eea
Thus we obtain
\bea
 \label{order}
&&
(2 \pi)^4 \delta^4\left(\sum p\right)
\left( \frac{g A_\mu}{\Lambda} \right)^A
\left( \frac{\pi}{f} \right)^B
\left( \frac{p}{\Lambda} \right)^C
f^2 \Lambda^2
 \nn\\
&\times& \left[ 
(4\pi)^{-2L} (f^{-1}\Lambda)^{2L}
\right] 
\left( \frac{g f}{\Lambda} \right)^{2I_A-2G} .
\eea
In order for the naive 
dimensional counting to be correct
when $I_A=0$,
the factor 
inside the square bracket should be of order one. 
Thus we set
\bea
 \label{applmd}
\Lambda \lesssim 4\pi f .
\eea

When $I_A \ne 0$,
there is a suppression factor $({gf}/{\Lambda})^{2I_A-2G}$.
($I_A \geq G$
since there cannot be a term with negative powers of $g$.)
The terms 
which do not preserve the 
weakly broken 
chiral symmetry (\ref{chiral})
are suppressed by this factor.

\subsubsection*{Taking into account the $N$ dependence}

In the above, we have neglected the
$N$ dependence,
related to the size of the gauge group. 
It can be taken into account by the following consideration
\cite{Soldate:1989fh}.
For vacuum diagrams which have no external legs
the number of the index loop is at most $L+1$,
which is the case when the diagram is planar.
Vacuum diagrams just contribute to the cosmological constant
which is only relevant when we consider the coupling to gravity
and we will not discuss it further in our effective field theory.
If there are external lines in the diagram,
they break at least one index loop compared with vacuum diagrams.
Thus the maximum number of the index loop is $L$.\footnote{%
For the estimation of the UV cut-off one should consider
the largest contributions to general operators in the model.
For a specific operator, 
the number of the index loops
in the Feynman diagrams which contribute
to that operator might be constrained to be smaller.
In that case, this term may be suppressed by the
inverse powers of $N$.
The double trace operator (\ref{highert}) is such an example.}
This modifies the $L$ dependent factor in
(\ref{order}) as
\bea
 \label{loopN}
\left[ 
(4\pi)^{-2L} (f^{-1}\Lambda)^{2L} (kN)^L
\right] .
\eea
Thus the appropriate UV cut-off is given by
\bea
 \label{ALambda}
\Lambda \approx \frac{4 \pi f}{\sqrt{kN}} .
\eea

\section{
The mass of the zero-modes in the one-loop effective potential}\label{Aoneloop}

Below we will estimate 
the mass of the zero-modes
in the effective potential at
the one-loop level (\ref{Veff2}):
\bea
 \label{aVeff2}
V_{1-loop} (u_0^I)
&=&
2 
\sum_{m_1,m_2} 
\sum_{i,j=1}^2
\int \frac{d^4 k}{(2\pi)^4} 
\ln \left(k^2 + m^2_{(m_1,m_2)(i,j)}(u_0^I) \right) .
\eea
Let us define
\bea
\zeta_{D^2} (s)
\equiv
\int \frac{d^4 k}{(2\pi)^4}
\sum_{m_1,m_2}\sum_{i,j=1}^2
\left(
k^2 + m_{(m_1,m_2)(i,j)}^2 (u_0^I)
\right)^{-s} .
\eea
Then, (\ref{aVeff2}) can be written as
\bea
V_{1-loop}(u_0^I) = - 2 \frac{d \zeta_{D^2}(s)}{ds} \Biggr|_{s=0} .
\eea
On the other hand, $\zeta_{D^2}$ can be rewritten as
\bea
\zeta_{D^2} (s) =
\frac{1}{\Gamma(s)} 
\int_0^\infty d\tau \,
\tau^{s-1}
\sum_{m_1,m_2}\sum_{i,j=1}^2
\exp\left[{-\tau \left( k^2 + m^2(u_0^I) \right)}\right],
\eea
where we have introduced 
a shorthand notation $m^2 (u_0^I)$ for $m_{(m_1,m_2)(i,j)} (u_0^I)$.
After performing the Gaussian integral, we obtain
\bea
\zeta_{D^2}
=
\frac{1}{(4\pi)^2}
\frac{1}{\Gamma(s)}
\sum_{m_1,m_2}
\sum_{i,j=1}^2
\int_0^\infty
d\tau \,
\tau^{s-3}
\exp\left[
-\tau m^2(u_0^i)
\right] .
\eea
Thus
\bea
\frac{d \zeta_{D^2}(s)}{ds}
=
\frac{1}{(4\pi)^2}
\sum_{m_1,m_2} \sum_{i,j=1}^2
\int_0^\infty d\tau \,
\tau^{s-3}
\left(
\frac{\ln \tau \, \Gamma(s)-\Gamma'(s)}{(\Gamma(s))^2}
\right)
\exp
\left[
-\tau m^2(u_0^I)
\right] .
\eea
Using
\bea
\lim_{s\rightarrow 0} \Gamma(s) = \frac{1}{s} + finite ,
\eea
we obtain
\bea
 \label{adzeta}
\frac{d \zeta_{D^2}(s)}{ds} \Biggr|_{s\rightarrow 0}
=
-
\frac{1}{2 (4\pi)^2 }
\sum_{m_1,m_2} \sum_{i,j=1}^2
\int_0^\infty
d\tau \,
\tau^{s-3}
\exp 
\left[ 
-\tau m^2 (u_0^I)
\right]
\Biggr |_{s \rightarrow 0}.
\eea
Now, recall (\ref{mass}):
\bea
 \label{amass}
m^2_{(m_1,m_2)(i,j)} (u_0^I) 
&\equiv&
\sum_{I =1,2}
\left(\frac{2}{a_I}\right)^2
\sin^2 \frac{1}{2} \left( m_I \theta + (u_i^I - u_j^I) \right) ,
\nn \\
&=&
\sum_{I=1,2}
\frac{2}{a_I^2}
\left(1 - \cos \left( m_I \theta + (u_i^I - u_j^I) \right) \right),
\eea
with
\bea
u_1^I = - u_2^I =  
\frac{1}{\sqrt{4N} f_I} u_0^I .
\eea
From (\ref{adzeta}) and (\ref{amass}), we obtain
\bea
 \label{aVeffz}
V_{1-loop} (u_0^I) 
&=&
-
\frac{1}{(4\pi)^2 }
\sum_{m_1,m_2} \sum_{i,j=1}^2
\int_0^\infty
d\tau \,
\tau^{s-3}
\exp \left[{-\tau \sum_{I=1,2} \frac{2}{a_I^2} }\right] \nn\\
&& \qquad \qquad
\times \exp\left[{\tau \sum_{I=1,2} \frac{2}{a_I^2} \cos (m_I \theta + u^I_{ij}) }\right]
\Biggr |_{s \rightarrow 0} ,
\eea
where
\bea
u^I_{ij} \equiv u^I_i - u^I_j .
\eea
We can safely set $s=0$ when
$N \geq 3$,
while for $N=2$ we have logarithmic divergence as a function of $s$.
Below we will consider the case $N \geq 3$ and set $s=0$.
In (\ref{aVeffz}) the sum over
$m_I$ $(I=1,2)$ run 
over integers in 
$ -\frac{N}{2} \leq m_I < \frac{N}{2}$.
Here we are considering the case $\theta = 2\pi / N$ and
due to the cancellation of the phases
only the terms proportional to
$\cos (\ell_I N (m_I \theta + u^I_{ij}))$
with integer $\ell_I$ survive in the sum over $m_I$.
Thus the net effect of the sum over $m_I$ with
$\theta = 2\pi / N$
is equivalent to the following Fourier transform:
\bea
\frac{1}{N} \sum_{m_I} f (\cos ( m_I \theta + u) )
=
\sum_{\ell_I = -\infty}^{\infty}
\left(
\frac{1}{\pi} \int_0^{2\pi} d\theta' \, 
f (\cos \theta')
\cos ( N \ell_I \theta') 
\right)
\cos ( N \ell_I u).
\eea
Using the identity for the
modified Bessel function $I_\nu (z)$ with integer $\nu$:
\bea
 \label{defmodB}
e^{z \cos \theta} =
I_0 (z) + 2 \sum_{\nu =1}^\infty I_\nu (z) \cos \nu \theta ,
\eea
(\ref{aVeffz}) can be rewritten as
\bea
V_{1-loop} (u_0^I) 
&=&
\frac{4 N^2}{(4\pi)^2} 
\int_0^\infty \frac{d\tau}{\tau^3} 
\exp \left[{-\tau \sum_I \frac{2}{a_I^2}}\right]
\sum_{\ell_1,\ell_2 = 0}^\infty
I_{N\ell_1}\left(\frac{2\tau}{a_1^2}\right)  
I_{N\ell_2}\left(\frac{2\tau}{a_2^2}\right) \nn\\
&& \qquad \qquad \times
\cos (N\ell_1 u^1_{ij})
\cos (N\ell_2 u^2_{ij}) .
\eea
When comparing the theories 
with different $N$,
we should fix the radii of the fuzzy torus (\ref{aftr})
and the $SU(k)$ gauge coupling as in (\ref{gscale}):
\bea
a_I N &=& 2\pi R_I : fixed, \\
g_{SU(k)} &\equiv& \frac{g}{\sqrt{N}} : fixed . \label{aagsuk}
\eea
Notice that this also fixes the scaling of $f$ 
through (\ref{agf}):
\bea
\frac{f}{\sqrt{N}} : fixed . \label{aafsuk}
\eea
We will express the calculations in terms of these parameters 
fixed for different $N$ below.

Using the
integral representation 
of the modified Bessel function
which follows from (\ref{defmodB}):
\bea
I_\nu (z)
=
\frac{1}{\pi}
\int_0^\pi \! d \theta' \, 
e^{z \cos \theta'} \cos \nu \theta'
\quad (\nu : integer) ,
\eea
and defining
\bea
V_{1-loop} (u_0^I)
=
4
\sum_{\ell_1=1}^\infty
\sum_{\ell_2=1}^\infty
V^{1-loop}_{(N\ell_1,N\ell_2)}
\cos (N\ell_1 u^1_{12})
\cos (N\ell_2 u^2_{12})
+ const. ,
\eea
we obtain
\bea
 \label{oneloopmass}
&& V^{1-loop}_{(N\ell_1,N\ell_2)} \nn\\
&=&
\frac{4 N^2}{(4\pi)^2} 
\int_0^\infty 
\frac{d\tau}{\tau^3} 
\prod_{I=1,2}
\frac{1}{\pi} \int_0^\pi d \theta_I 
\exp \left[{-\tau \frac{2N^2}{(2\pi R_I)^2}(1-\cos \theta_I ) }\right]
\frac{e^{i N \ell_I \theta_I} + e^{-i N \ell_I \theta_I}}{2}
\nn\\
&=&
\frac{N^2}{(4\pi)^2} \int_0^\infty 
\frac{d\tau}{\tau^3} 
\prod_{I=1,2}
\frac{1}{\pi} \int_0^\pi d \theta_I 
\Biggl(
\exp
\left[-\tau \frac{1}{(2\pi R_I)^2} 
 \left(
N^2 \theta_I^2 + i \frac{(2\pi R_I)^2}{\tau} \ell_I N \theta_I
 \right)
\right] \nn\\
&& \qquad \qquad \qquad \qquad +  
\exp
\left[-\tau \frac{1}{(2\pi R_I)^2} 
 \left(
N \theta_I^2 - i \frac{(2\pi R_I)^2}{\tau} \ell_I N \theta_I
 \right) 
\right]
+ {\cal O}(N^{-2}) 
\nn\\
&=&
\frac{1}{(4\pi)^2} 
\int_0^\infty 
\frac{d\tau}{\tau^3} 
\prod_{I=1,2}
\frac{1}{\pi} \int_0^{N\pi} d \ttheta_I 
\Biggl(
\exp
\left[ - \frac{\tau}{(2\pi R_I)^2} 
 \left(
 \ttheta_I^2 + i \frac{(2\pi R_I)^2}{\tau} \ell_I  \ttheta_I 
 \right)
\right] \nn \\
&& \qquad \qquad \qquad \qquad  +  
\exp
\left[- \frac{\tau}{(2\pi R_I)^2} 
 \left(
  \ttheta_I^2 - i \frac{(2\pi R_I)^2}{\tau} \ell_I  \ttheta_I
 \right)
\right]
\Biggr)
+ {\cal O}(N^{-2}) 
\nn\\
&& \qquad (\ttheta = N \theta)
\nn \\
&=&
\frac{1}{(4\pi)^2} 
\int_0^\infty 
\frac{d\tau}{\tau^3} 
\frac{4}{(2\pi)^2}
\left(
\prod_{I=1,2}
\sqrt{\frac{\pi (2\pi R_I)^2}{\tau}}
e^{-\frac{(2\pi R_I)^2 \ell_I^2}{4\tau}}
\right)
+ {\cal O}(N^{-2}) +  {\cal O}\left( \frac{1}{N} e^{- (\pi N)^2} \right) 
\nn\\
&=&
\frac{1}{(4\pi)^2} 
\int_0^\infty d\ttau 
\ttau^{2}
\frac{1}{\pi}
\left(
\prod_{I=1,2}
2\pi R_I e^{-\frac{(2\pi R_I)^2 \ell_I^2}{4} \ttau}
\right)
+ {\cal O}(N^{-2}) +  {\cal O}\left( \frac{1}{N}e^{- (\pi N)^{2}}\right) 
\nn\\
&& \qquad \left( \ttau = \frac{1}{\tau} \right)
\nn \\
&=&
\frac{1}{(4\pi)^2} 
\Gamma(3)
\frac{1}{\pi} (2\pi R_1) (2\pi R_2) 
\left( 
\frac{4}{(2\pi R_1)^2 \ell_1^2 + (2 \pi R_2)^2 \ell_2^2}
\right)^{3} 
+ {\cal O}(N^{-2}) + {\cal O}\left( \frac{1}{N}e^{- (\pi N)^{2}}\right). \nn\\
&& 
\eea
The leading term 
in the large $N$ expansion
coincides with the one
in the gauge-Higgs unification in
the ordinary torus extra dimensions
\cite{Hosotani:1983xw,Hosotani:1988bm,Davies:1988wt,Hatanaka:1998yp}. 
Noticing that in terms of the canonically normalized field $u_0^I$,
\bea
N u_{12}^I =
N \frac{1}{\sqrt{4N} f_I} 2 u_0^I
=
g_{SU(k)} (2\pi R_I) u_0^I ,
\eea
the mass $m_0$ of the zero-modes can be read off from
(\ref{oneloopmass}) and is of order
\bea
 \label{zmass}
m_0^2
\approx
\frac{g_{SU(k)}^2}{16\pi^2}
\frac{1}{R^2} ,
\eea
where the four-dimensional effective $SU(k)$ gauge coupling 
$g_{SU(k)}$ is given in (\ref{aagsuk}) 
and
as before we have assumed $R_1 \approx R_2 \approx R$.
This is as expected since
$1/R$ is the scale where 
the effect of the new physics appears,
and
$g_{SU(k)}^2 / {16\pi^2}$
is the one-loop factor.

${\cal O}(N^{-2})$ etc. in (\ref{oneloopmass})
refers to the relative magnitude compared with the leading term.
Thus even when $N$ is small, it gives a correction
at most of the same order.
Thus (\ref{zmass}) is still a valid order estimate
even when $N$ is small.

\bibliography{fuzzXref}
\bibliographystyle{utphys}

\end{document}